\documentclass[sigconf,nonacm]{acmart}

\AtBeginDocument{%
  }

\begin{document}

\title{MQTT Across a Raspberry Pi 5 IoT Network Utilizing Quantum-resistant Signature Algorithms}

\author{Ray Feingold}
\email{r.feingold@vikes.csuohio.edu}
\affiliation{%
  \institution{Cleveland State University}
  \city{Cleveland}
  \state{Ohio}
  \country{USA}
}

\author{Chansu Yu}
\email{c.yu91@csuohio.edu}
\affiliation{%
  \institution{Cleveland State University}
  \city{Cleveland}
  \state{Ohio}
  \country{USA}
}
\begin{abstract}
The rapid expansion of the Internet of Things (IoT) has introduced millions of resource-constrained devices into critical infrastructures, consumer environments, and industrial systems. These devices rely on lightweight communication protocols such as MQTT to support low-power, intermittent, and bandwidth-limited operation. However, common TLS algorithms used to secure MQTT communications are vulnerable to quantum attacks made feasible by Shor’s algorithm. As a result, IoT infrastructures must evaluate and adopt post-quantum cryptographic (PQC) methods capable of providing long-term resilience.

This report investigates the implementation of PQC algorithms within an MQTT-based IoT networks using three Raspberry Pis. Specifically, it integrates the FALCON digital signature scheme, one of NIST’s selected post-quantum signature algorithms, to maintain message authenticity and integrity across resource-constrained MQTT clients and brokers. By measuring system performance, the research characterizes the practical trade-offs of deploying lattice-based PQC on lightweight hardware. 
\end{abstract}

\keywords{}


\maketitle

\section{Introduction}
With the proliferation of Wi-Fi over the past few decades, the adoption of everyday technologies into Wi-Fi-enabled Internet of Things (IoT) devices continues to increase dramatically. Through this, the way information is collected, transmitted, stored, and acted upon in modern digital infrastructures has changed fundamentally. Sensors, embedded systems, and everyday smart appliances now routinely interact with network services to provide real-time monitoring, data analytics, and automation controls. As IoT devices continue to scale in number and importance, their reliance on secure and efficient communication protocols has become a critical component of their design. Lightweight protocols such as MQTT help enable these processes by offering a simple, low-overhead messaging model designed to function over resource-constrained devices and unreliable networks. This allows distributed IoT systems to efficiently exchange data over unreliable or high-latency networks, which makes it well suited for bandwith-constrained and intermittently connected environments, reflecting MQTT's core design goal \cite{hivemq_mqtt_essentials}. 

Emerging capabilities in quantum computing directly threatens the long-term security of IoT networks. Algorithms like Shor’s algorithm pose a direct risk to widely deployed cryptographic algorithms – particularly RSA and elliptic curve schemes – which interop with most network authentication methods, key exchanges, and certificate infrastructures [1]. Due to many IoT devices remaining in service for years to even decades, some are expected to be operational as the cryptographic capabilities of quantum computers are realized [5]. This long device lifespan coupled with delays in update cycles for deployed systems, makes IoT infrastructures a vulnerable domain in the transition to post-quantum security.

Post-quantum cryptography proposes new cryptographic primitives resistant to attacks by both classical and quantum computers. Despite advancements in algorithmic capabilities, PQC integration into an IoT ecosystem remains a challenge. Algorithms often require larger keys and signatures, introducing substantial computational overhead and memory allocation concerns. With these issues under consideration, the National Institute of Standards and Technology (NIST) has selected several candidate algorithms for standardization in recent years. Included in this is FALCON, a lattice-based digital signature scheme designed for compactness and efficiency, with key sizes smaller than other NIST selected PQC signature schemes such as CRYSTALS-Dilithium and SPHINCS+ \cite{nist_pqc}. This proves desirable when implementing such methods on resource-constrained IoT devices, allowing for efficiency increases with PQC algorithm usage.

This paper explores the integration of PQC algorithms in an MQTT-based IoT environment using three Raspberry Pi 5s to model an IoT publish-subscribe model, routing all communications through a central message broker. By adopting the FALCON digital signature algorithm, this study examines the effect PQC has upon communication latency, device performance, and overall system behavior. Overall, the goal is to evaluate PQC-enabled MQTT deployment and its practicality for resource-constrained IoT devices, identifying strategies for achieving cryptographic agility during the transition into a post-quantum computing world. The full implementation, including setup scripts and configuration files, 
is publicly available in our GitHub repository~\cite{oqs_demos_mosquitto}.
\subsection{Related Work}
This paper modifies and builds upon the work found in the Open Quantum Safe (OQS) demonstrations for the MQTT software package \textit{mosquitto}. While their repository and associated demo adequately creates an MQTT-enabled communication system via pre-built Docker images, it does not provide a practical implementation of its usage across an IoT network. This is due to the fact that Docker containers abstract away the physical hardware layer, which removes real device provisioning, GPIO interfacing, certificate distribution across nodes, and the networking constraints that are inherent to embedded IoT endpoints. To address this gap, our goal is to construct an easily accessible starting point for those desiring a practical, quantum-safe, and MQTT-enabled IoT system deployable on real hardware. 
\subsection{Paper Organization and Contributions}
The organization and contributions of this work is as follows:
\begin{itemize}
    \item We overview and analyze key components of the developed Raspberry Pi MQTT IoT system (Section 2).
    \item We present a high-level overview of the IoT system architecture created (Section 3).
    \item We provide a guide on system setup and results (Sections 4, 5).
\end{itemize}
\section{Background}
This section provides a high-level overview for the fundamental infrastructure components for the system created. Similarly, the potential benefits and necessities for PQC-enabled IoT devices are further explored. 
\subsection{The Internet of Things (IoT)}
The Internet of Things (IoT) refers to interconnected systems of sensors, appliances, and embedded devices that incorporate networking and computational capabilities. Examples include ‘smart’ devices such as fridges, thermostats, doorbell cameras, etc. These devices allow physical environments to be monitored, controlled, analyzed, and automated. IoT architectures typically follow a layered model consisting of \cite{salih_iot_survey_2022}:
\begin{itemize}
    \item \textbf{Device / sensing layer}: physical sensors and actuators that generate or respond to environmental data.
    \item \textbf{Network layer}: connectivity protocols such as MQTT, CoAP, Wi-Fi, Bluetooth, Zigbee, or cellular.
    \item \textbf{Middleware layer}: gateways, edge processers.
    \item \textbf{Application layer}: cloud services, dashboards, analytics, end-user interactivity.
\end{itemize}
In an IoT system, data flows from endpoints through gateways or brokers to higher-level applications, allowing devices with minimal computing power to participate in broader distributed systems [7]. This layered design approach separates functionalities such that established infrastructures are scalable. Though this can introduce security and interoperability challenges.

IoT devices vary dramatically in memory, computational power, energy availability, and networking performance. Some devices operate on limited batteries and remain dormant for extended periods of time; others serve as continuously running gateways with more computational resources and capabilities. These constraints directly influence protocol selection, cryptographic feasibility, and update strategies. Lightweight communication and compact security primitives prove essential for avoiding degrading device performance, increasing latency, or exhausting limited stored energy [5]. 
    
Security within IoT systems remains multifaceted. Devices may exist actively deployed in the field for decades, increasing the importance of secure boot, firmware updates, and long-term cryptographic resilience. Physical access risks, heterogeneous device management, and large-scale deployments further complicate key distribution, credential rotation, and trust establishment. Given the increasing likelihood of future quantum-capable adversaries, IoT architectures must consider cryptographic agility and design assumptions that remain valid for many years, essentially future-proofing the infrastructure.
\subsection{Message Queuing Telemetry Transport (MQTT)}
Message Queuing Telemetry Transport (MQTT) is a lightweight publish-subscribe messaging protocol designed to accommodate low-power and low-bandwidth environments [3]. Its architecture revolves around a central broker that manages communications between publisher endpoints, which send messages, and subscriber endpoints that receive messages through named topic channels. This decoupling of clients simplifies IoT system design, enabling devices to communicate without requiring direct connections with one another.

MQTT operates over TCP and provides three quality of service (QoS) levels \cite{hivemq_mqtt_essentials}:
\begin{itemize}
    \item \textbf{QoS 0}: One delivery at most.
    \item \textbf{QoS 1}: One delivery at least.
    \item \textbf{QoS 2}: One delivery exactly.
\end{itemize}
These options allow for the balance of reliability and system overheads depending upon application requirements. (citation / add more in this paragraph)

MQTT’s simplicity and low overhead make it suitable for IoT environments, but the protocol itself does not enforce built-in security mechanisms. TLS is commonly used to secure transport layer communications, while authentication is typically implemented via broker-managed credentials or external systems. For scenarios requiring end-to-end confidentiality or integrity – especially across untrusted brokers – payload-level encryption becomes necessary. PQC-based protections, such as the FALCON signature scheme, can be utilized across MQTT functionalities to enhance message authenticity and long-term cryptographic resilience [5].
    
The addition of PQC to MQTT introduces its own challenges, however. Larger signatures may increase bandwidth consumption, more compute-intensive signing and verification may add latency, and constrained hardware may struggle with key generation or repeated cryptographic operations [5]. Understanding these trade-offs is central to evaluating the feasibility of PQC-enhanced MQTT in real-world IoT deployments. 	
\subsection{Post quantum cryptography (PQC)}
Post-quantum cryptography refers to algorithms designed to remain secure despite quantum computer advancements. The principal motivation for PQC arises from Shor’s algorithm’s efficient solving of mathematical operations that underlie RSA, Diffie-Hellman, and elliptic-curve cryptography [1]. As these methods form the main cryptographic components of modern public-key infrastructures, any compromises would undermine countless secure digital communications.

To address this issue, NIST ran a multi-year-long effort to evaluate potential quantum-resistant algorithms. The candidates span several algorithmic families, including lattice-based, code-based, hash-based, and multivariate-polynomial schemes. These algorithms are based upon mathematical problems difficult for even large-scale quantum computers to solve efficiently. Overall, there is a desire to provide alternatives for all major cryptographic functions, like key exchanges and digital signatures, without impacting system performance or longevity.
    
Although PQC algorithms strengthen long-term security, this is often accompanied by practical drawbacks: larger public keys, larger signatures, increased computational overhead, or more complex implementation requirements. These issues pose significant challenges for IoT devices, which frequently operate under specified performance and memory constraints. As systems transition into quantum computing-resistant architectures, adopting hybrid or gradually phased-in approaches to infrastructure updates may be necessary to maintain compatibility while still addressing security concerns.
\subsection{FALCON signature scheme}
Included among the PQC digital signature schemes selected by NIST for standardization is \textit{Fast Fourier Lattice-Based Compact Signatures over NRTU}, or FALCON.  Utilizing the Gentry–Peikert–Vaikuntanathan (GPV) framework, this algorithm integrates NTRU lattices coupled with fast Fourier-based sampling methods to generate signatures capable of resisting large-scale quantum computing decryption attempts [1] [4] [5] [6] [8]. Furthermore, one of FALCON’s most notable features is its compactness, featuring small key and signature sizes. This proves desirable for IoT devices limited to certain power and bandwidth specifications, allowing for secure communication without compromising overall performance [1]. In this way, MQTT IoT infrastructures can be preemptively configured to employ quantum-safe techniques across end devices.

Underlying FALCON is lattice-based cryptography, a class of cryptographic methods in which security measures rely upon the computational complexity in solving mathematical problems with high-dimensional geometric lattices. In this context, a lattice can be understood as a regularly structured grid of points in many dimensions, where finding a point within the approximate vicinity of a given target vector remains unlikely even for proposed quantum computer decryption capabilities [8]. The GPV framework produces digital signatures by sampling short lattice vectors [1]. This obscures the private key from external observation while allowing for public key verification. Together, the NTRU lattice structure and GPV signing method enable FALCON to generate compact, secure signatures that remain resistant to both classical and quantum attacks. 

\subsection{Associated Security Libraries with PQC MQTT Capabilities}
To accomplish secure MQTT communications via PQC-integrated digital signature algorithms, a few related software libraries must be used to fulfill implementation requirements. Due to PQC algorithms suffering slow adoption, common network libraries, like \textit{openssl}, must be compiled from source manually with encapsulated PQC algorithmic functionalities. Without doing so, the system may face software-related technical difficulties stemming from mismatched versions or other similar issues. In this IoT system, the \textit{liboqs}, \textit{oqs-provider}, \textit{mosquitto}, and \textit{openssl} package libraries shall be used primarily to allow for PQC interoperability.  Below is more information on each library:
\begin{itemize}
    \item \textbf{openssl}: A commercially used open-source command-line tool created to secure TLS/SSL communications. Often, it is used for SSL/TLS certificate management, such as the generation of private and public keys, certificate signing requests, and generalized X.509 certificates. [10]
    \item \textbf{liboqs}: An open-source C library designed to provide quantum-safe algorithms for software applications or packages. Within it, several algorithms, including ones standardized by NIST or otherwise, are supplied, as well as API wrappers for other programming languages. This library is maintained by the Open Quantum Safe (OQS) project, an open-source collaboration team dedicated to preparing commonly used encryption-based libraries for post-quantum computing decryption attacks. [9]
    \item \textbf{oqsprovider}: A tangentially created OQS library created to provide a single, shared library for usage with \textit{openssl}. It contributes several quantum-safe algorithms enabled for the \textit{openssl} command-line package, including FALCON. [11]
    \item \textbf{mosquitto}: An open-source C library capable of implementing the MQTT protocol across IoT systems. When used in tandem with the aforementioned OQS libraries, communication between end devices can be PQC-enabled and secured. [12]
\end{itemize}
\section{Design}
In this section, we provide a high-level overview of the system architecture implemented. It consists of three Raspberry Pi 5 endpoints all operating within the same local network, forming a small-scale representation of common IoT environments. One Raspberry Pi is connected to a motion-detection sensor, capturing movement events and communicating it to the specified topic channel \textit{motion-sensor}. The remaining two Raspberry Pis act as the broker and the subscriber, respectively, creating a complete publish-subscribe model enabled for real-time data collection. 
\subsection{Motion sensor system}
The motion sensor system consists of three Raspberry Pi 5 endpoints arranged in a standard MQTT publish-subscribe model as well as a simple sensor circuit. As one device receives environmental motion detection data from an attached sensor, it publishes relevant data to a designated channel subscribed to by another endpoint. Between these two workstations lies a broker, facilitating communication between publishers and subscribers by routing incoming sensor data to \textit{motion-sensor}. Figure 1 is a diagram outlining the circuit attached to the publisher Raspberry Pi.

\begin{figure}[h]
    \centering
    \includegraphics[width=\linewidth]{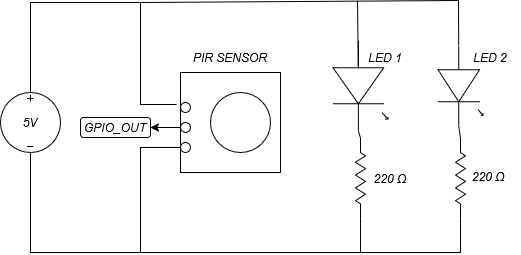} 
    \caption{Motion detecting circuit attached to publisher Raspberry Pi}
    \label{fig:circuit}
\end{figure}

As motion is detected by the sensor, data travels from the publisher to the designated motion-sensor topic channel for bidirectional communications to the subscriber device. Any device subscribed to the same channel will receive this information in near real time, including the configured subscriber Raspberry Pi. This flow models a typical IoT pipeline, where data originates from physical hardware and propagates to interested endpoints using a messaging protocol. Similarly, this system allows multiple subscribers to be added without requiring changes to the motion sensor circuit itself.
\subsection{Publisher}
Physically connected to the motion detection circuit is our designated Raspberry Pi publisher device responsible for generating MQTT messages. As the sensor identifies movement within its local environment, the publisher translates the sensor data into a digital payload. This payload is then transmitted to the \textit{motion-sensor} topic channel. Furthermore, and for testing purposes, a heartbeat message is sent every 60 seconds to only the broker Raspberry Pi, verifying the status and reliability of the connected hardware. Together, the sensor data and heartbeat messages provide both functional data and system health information.
\subsection{Broker}
One Raspberry Pi 5 acts as a broker, facilitating communication between the publisher attached to the motion detection system and relevant subscriber endpoints. In MQTT, a broker device serves as the central point to which messages are sent and received in an IoT system \cite{hivemq_mqtt_essentials}. Moreover, in our model, messages from the publisher and the attached motion detection system are received by the broker and then distributed across the message channel to the subscribed Raspberry Pi 5. While the only channel used by our system is \textit{motion-sensor}, an MQTT broker can filter and forward ingested data to the appropriately subscribed clients efficiently.
\subsection{Subscriber}
The remaining Raspberry Pi 5 operates as an endpoint subscribed to the topic channel \textit{motion-sensor}. As messages propagate from the motion detection circuit to the attached publisher endpoint, the broker communicates these accordingly to the subscriber. These messages can then be displayed, logged, or further processed depending on system configurations and requirements. Furthermore, the subscriber endpoint represents the consumer of IoT data, like other sensing-based deployments. By separating the publisher and subscriber roles, the architecture mirrors how data is typically distributed across IoT networks, especially within the context of an MQTT-enabled infrastructure.
\subsection{High-level Overview of Architecture}
In summary, our designed IoT infrastructure consists of three Raspberry Pi 5s, with one possessing a physically attached motion detection sensor and associated circuit. Figure 2 displays the overall layout of the architecture.
\begin{figure}[h]
    \centering
    \includegraphics[width=\linewidth]{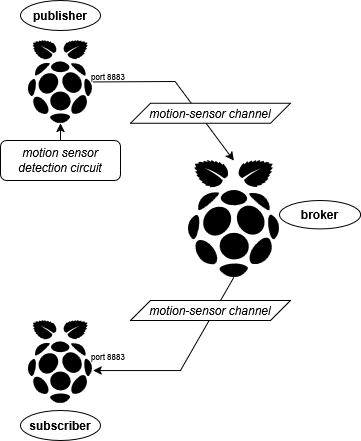} 
    \caption{High-level architecture of MQTT IoT system}
    \label{fig:diagram}
\end{figure}
\section{Setup}
This paper comes with a supplementary GitHub repository containing the code for setting up the IoT network. The repository is publicly available at: \textit{https://github.com/rayfcsu/pqc-mqtt}. All components required for deploying this system are within it. 

The repository extends the OQS \textit{mosquitto} demonstration by integrating a functional IoT application, like a motion-detection sensor network, secured entirely using PQC algorithms [13]. Instead of providing minimal TLS/SSL testbed, this implementation deploys a complete, operational IoT workflow consisting of sensing, message publishing, brokering, and subscribing all protected by quantum-safe techniques. Similarly, the system utilizes the FALCON-1024 digital signature algorithm for certificate-based authentication, thereby ensuring cryptographic resilience against large-scale quantum computing decryption efforts.

In this section, we provide a brief overview and guide on how to set up the designed PQC and MQTT-enabled Raspberry Pi IoT system using the listed GitHub repository and physical circuit elements.
\subsection{Assembling the motion sensor system}
To construct the motion-sensing IoT system, the following hardware components are required:
\begin{itemize}
    \item Three Raspberry Pi 5 units
    \item One breadboard
    \item One HC-SR501 PIR motion sensor
    \item Three female-to-female jumper wires
    \item Three female-to-male jumper wires
    \item Two 220 ohm or 300 ohm resistors
    \item Two LEDs
\end{itemize}
In our circuit, two LEDs are utilized with one illuminating in the presence of motion detection, and the other remaining continuously lit to display system validity. The PIR sensor and LEDs are connected to the publisher Raspberry Pi according to the wiring diagram shown in Figure 2. Another diagram can be found in reference [2] as well. The default pin assignments are:
\begin{itemize}
    \item PIR sensor: GPIO14 (BCM14, or physical pin 8)
    \item Status LED: GPIO21 (BCM21, or physical pin 40)
    \item Detection LED: GPIO20 (BCM20, or physical pin 38)
\end{itemize}
These pin mappings are hard-coded into the publisher startup script and must be preserved unless the script is modified accordingly. A pin assignment chart for the GPIO interface of a Raspberry Pi 5 can be found online if there is a desire to change these.

All Raspberry Pi devices must have Secure Shell (SSH) enabled and be assigned static or known local IP addresses for usage later. They must also be able to connect to the internet for the purpose of downloading the repository and required packages during initial setup.
\subsection{Downloading the GitHub repository}
On each Raspberry Pi, the project repository is obtained by cloning it from GitHub. This repository contains all scripts required to compile the PQC libraries, configure the MQTT broker, generate the required certificates, and launch the publisher and subscriber nodes. 
\subsection{Configuring the environment setup}
Before deploying the MQTT system, all PQC-enabled dependencies must be compiled and installed. This can be achieved by executing the environment setup script \textit{pqc-mqtt-env-setup-sh}, and instructions on how to run it are listed within the GitHub repository. This script installs all required system packages and build tools, downloads and compiles the \textit{liboqs} and \textit{oqsprovider} libraries, \textit{openssl}, and a post-quantum-enabled build of the \textit{mosquitto} MQTT service, configuring the necessary shared library paths and environment variables simultaneously. In addition, it creates the \textit{pqc-mqtt} working directory used to store runtime files and device certificates. Upon completion of this step, each Raspberry Pi is fully prepared to support PQC-enabled communication. 

Also included within the repository is a generalized cleanup script, \textit{pqc-mqtt-env-cleanup.sh}, meant to remove all installed binaries and packages and provide a fresh environment as needed.
\subsection{Executing the system scripts}
Within the given repository, there are three startup shell scripts provided. These are used in initializing the MQTT connections of the Raspberry Pi nodes and all end in \textit{-start} in the file names.
\subsubsection{Broker and certificate authority initialization}
The MQTT broker and the system’s certificate authority are initialized on the designated broker node by executing the \textit{broker-start.sh} script. During execution, the script prompts the user to provide the broker’s IP address as well as the IP addresses and SSH usernames of the publisher and subscriber nodes. These values must be known and exported as variables to a user-created file named \textit{pqc-env.sh}. Using this information, the script generates a FALCON-1024 certificate authority key pair, creates and signs a broker certificate, and securely distributes the CA certificate and key to the publisher and subscriber. It then constructs a \textit{mosquitto} configuration file that enables the PQC-enabled TLS 1.3 connection, launching the \textit{mosquitto} broker on TCP port 8883 when it concludes. At this point, a fully authenticated and quantum-safe MQTT broker is now available for device connections. 
\subsubsection{Publisher and motion sensor initialization}
On the Raspberry Pi connected to the PIR motion sensor, the publisher service is started by executing the publisher-start.sh script. After prompting the user for the broker and publisher IP addresses, the script generates a FALCON-1024 certificate for the publisher signed by the system certificate authority. Then, it initializes the required GPIO pins for the PIR sensor and the status and detection LEDs before establishing a connection to the MQTT broker. Once connected, the publisher continuously monitors the sensor and publishes all detection events to the motion-sensor topic channel, transmitting periodic heartbeat messages to the status channel. In this way, the publisher operates as a secure IoT sensing node that reports environmental activity in near real time.
\subsubsection{Subscriber initialization}
For the subscriber device, the MQTT client is launched by utilizing the \textit{subscriber-start.sh} script. This script requests the broker and subscriber IP addresses, generating a FALCON-1024 certificate for use with its connection to the broker. After successfully connecting, the subscriber registers to the motion-sensor topic channel and displays all received motion events in near real time, thereby completing the secure end-to-end IoT communication pipeline.
\section{Testing}
To test the usage and validity of the designed MQTT-enabled IoT system, we compare the key and certificate generation portion of the \textit{publisher-start.sh} and \textit{subscriber-start.sh} scripts. Because testing the \textit{broker-start.sh} script is more difficult due to how it runs \textit{mosquitto}, it is assumed that its experimental results are similar to the publisher and subscriber Raspberry Pi nodes.  
\subsection{Setup}
For testing the time complexity of the cert generation, two small scripts are used: \textit{run-publisher-tests.sh} and \textit{run-subscriber-tests.sh}. These iterate through the \textit{publisher-start.sh} and \textit{subscriber-start.sh} scripts, respectively, 25 times for the RSA-2048 and FALCON-1024 digital signature algorithms. Each run outputs the results in nanoseconds to a CSV file named \textit{results.csv} before continuing on to the next iteration. In total, 25 runs for each algorithm were completed on both the subscriber and publisher endpoints. 
\subsection{Performance results}
Below are the results of running the aforementioned testing scripts:
\begin{figure}[h]
    \centering
    \includegraphics[width=\linewidth]{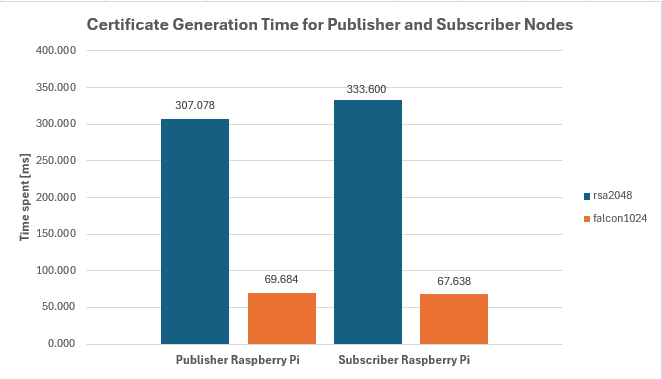} 
    \caption{Comparison of certificate generation time by certificate type on publisher and subscriber Raspberry Pi nodes}
    \label{fig:pqc-test-results}
\end{figure}
\subsection{Comments}
The results shown in Figure 3 indicate a significant difference in certificate generation time between the classical RSA-2048 and quantum-safe FALCON-1024 signature scheme. On both the publisher and subscriber Raspberry Pi nodes, FALCON-1024 required substantially less time to generate certificates, completing the process in approximately 68-70 milliseconds on average. In contrast, the RSA-2048 signature scheme required over 300 milliseconds on each device. This reduction in generation time is notable given the common perception that PQC algorithms are more computationally expensive than classical alternatives, which can be concerning resource-constrained embedded devices. This perception holds for many PQC schemes, which feature larger keys and more intensive operations. However, RSA's certificate generation cost is dominated by expensive large integer primality testing and modular exponentiation, which scales poorly with key size. FALCON-1024, on the other hand, relies on polynomial arithmetic over NTRU lattices and fast Fourier transforms. Operations that are more efficient at equivalent security levels, which makes it a compelling exception to the general assumption of PQC overhead. Therefore, the compact structure and efficient sampling methods used by FALCON allow it to outperform RSA for certificate creation.

These findings have critical implications for the deployment of PQC in IoT environments. Certificate generation remains imperative to securing MQTT systems as it directly impacts device provisioning, authentication, and recovery procedures. Faster certificate creation enables embedded hardware to communicate among other devices with limited effects on latency, bandwidth, and overall performance. Furthermore, the results observed between the publisher and subscriber nodes also suggest that FALCON behaves consistently across devices, reinforcing its suitability for distributed systems. Overall, the experimental results demonstrate that post-quantum signatures not only provide long-term security benefits, but offers additional practical performance advantages, producing a viable option for securing current and future IoT infrastructures.
\section{Conclusion}
In this paper, we demonstrate that PQC mechanisms can be effectively integrated into a lightweight IoT messaging architecture without disrupting overall system functionality. By deploying the FALCON digital signature scheme within an MQTT-based network of Raspberry Pi 5 endpoints, this work shows that quantum-resistant authentication, certificate handling, and message integrity can be achieved in an IoT environment. Sensor data was successfully collected and delivered to a subscriber using PQC-enabled communications, confirming that lattice-based cryptography can operate within the constraints of embedded hardware. The use of modular cryptographic libraries such as OpenSSL, liboqs, oqsprovider, and Mosquitto enabled PQC support without modifying application-level MQTT logic, showing how quantum-safe implementations can be achieved through standardized software. Although PQC algorithms are generally considered to introduce higher computational costs than classical schemes, FALCON demonstrated competitive and in some cases superior performance, while the system remained responsive and stable throughout deployment.

Broadly, this work emphasizes the importance of preparing IoT infrastructures for post-quantum computing decryption techniques before quantum computers are physically realized. Because many IoT devices are expected to remain deployed for extended periods of time, delaying cryptographic upgrades would expose them to long-term security risks. By validating FALCON-based MQTT communication in a functional sensing network, this paper provides evidence that PQC can be deployed proactively rather than retroactively. As IoT continues to expand into critical domains, integrating PQC into its communication layers will be essential for ensuring enduring quantum-safe security and resilience.

\subsection{Future Work}
While FALCON has been proven to be effective for PQC-enabled MQTT deployment, it introduces potential timing leaks and side-channel vulnerabilities by its reliance on high-precision Gaussian sampling and floating-point operations [1], [4], [6], and [8]. A promising direction for future work is the exploration of SOLMAE \cite{solmae}, a lattice-based signature scheme that is similar to FALCON in compactness and efficiency, but is designed with explicit resilience against side-channel attacks. Evaluating SOLMAE within the same Raspberry Pi IoT testbed would provide a direct comparison and can further strengthen the cryptographic agility of the proposed system.

\appendix

\end{document}